\documentclass[pdftex,12pt,a4paper]{article}
\usepackage{amsmath}  
\usepackage{amssymb}
\usepackage{amsfonts}
\usepackage{mathptmx}
\usepackage{latexsym}
\usepackage{setspace}
\usepackage{graphicx,subfigure}
\usepackage{color}
\usepackage{enumerate}
\usepackage{enumitem}
\usepackage[ocgcolorlinks]{hyperref}
\hypersetup{
  citecolor=blue,
 linkcolor=blue,
 backref=true,pagebackref=true
 }
\usepackage{fancyhdr} 
\begin{document}

\title{Grain boundary ferromagnetism in vanadium-doped In$_2$O$_3$ thin  films}
\date{}
\maketitle

\vspace{-1in}

{\centering {\bf Qi Feng\footnote{Present address: SKLSM, Institute of Semiconductors, Chinese Academy of Sciences,  P. O. Box 912, Beijing  100083, People's Republic of China}, Harry J Blythe, A Mark Fox, and Gillian A Gehring\footnote{g.gehring@sheffield.ac.uk}} \\{\it Department of Physics and Astronomy, The University of Sheffield, S3 7RH, UK} \\{\bf Xiu-Fang Qin, and Xiao-Hong Xu}\\{\it School of Chemistry and Materials Science, Shanxi Normal University, Linfen 041004, People's Republic of China}\\{\bf  Steve M Heald}\\{\it Advanced Photon Source, Argonne National Laboratory, Argonne, Illinois, 60439, USA}}


\begin{abstract}

Room temperature ferromagnetism  was observed in In$_2$O$_3$ thin films doped with 5 at.\% vanadium, prepared  by pulsed laser deposition  at substrate temperatures ranging from  300 to 600 $\,^{\circ}{\rm C}$. X-ray absorption fine structure  measurement indicated that vanadium was  substitutionally dissolved   in the In$_2$O$_3$ host lattice, thus excluding the existence of secondary phases of vanadium compounds. Magnetic measurements based on SQUID magnetometry and magnetic circular dichroism  confirm that the  magnetism is at grain boundaries and also in the grains. The overall magnetization originates from the competing effects between grains and  grain boundaries.


\end{abstract}
\newpage
\section{Introduction}

Dilute magnetic oxides (DMOs) have been identified as one of the most attractive materials for future spintronics applications \cite{Wolf.Sci.2001, Zutic.2004, Coey.Nat.Mat.2005,   Bader.Annu.Rev.Cond.Mat.Phys.2010}. However the field has also been very controversial.  In some cases the magnetism has been attributed to ferromagnetic nano-inclusions \cite{Kim.APL.2002, Heald.PRB.2009} although there have been a number of  cases where the magnetic properties were identified in samples without any impurity phases leading to the conclusion that a `non-trivial defect-related' ferromagnetic phase exists in some oxides \cite{Coey.COSSM.2006}. The possible explanations of oxide magnetism were further limited by the X-ray magnetic circular dichroism  experiment which showed that Co ions in ZnCoO were themselves paramagnetic  \cite{Tietze.NJP.2008}; this startling result has been confirmed by further investigations of other doped magnetic semiconductors \cite{Barla.PRB.2007, Ney.NJP.2010, Hakimi.PRB.2010, Sakai.JMMM.2012}.  In addition, large magnetic moments have also been observed in undoped oxides \cite{Hong.PSS.2007}.  This has meant that  acceptable models should now include those where the magnetism is associated with electrons in states other than on the transition-metal (TM) ions themselves.

Another feature of TM-doped oxide materials is that ferromagnetism (FM) is absent in epitaxial films and in the bulk, this was quantified by Straumal \emph{et al} \cite{Straumal.PRB.2009, Straumal.JETP.2013}  who proposed a model of grain boundary (GB) FM  in pure and doped ZnO.  They found that FM only occurred if the grains were sufficiently small $g<g_{c}$ where $g_{c}$, the critical grain size, was increased if Mn was doped into ZnO, a similar effect has been seen in Mn-doped In$_2$O$_3$ \cite{Qi.APL.Mat.2013}. Coey \cite{Coey.NJP.2010} proposed that the FM  originates from a narrow spin split defect band whose population is controlled by doping TM ions.  There are two obvious candidates for such a band: it might arise from states localized at the grain boundaries as suggested by the work of Straumal \emph{et al} \cite{Straumal.PRB.2009, Straumal.JETP.2013} or it might be the  narrow band originating from states at oxygen vacancies as had been proposed earlier for both ZnO and In$_2$O$_3$ \cite{Lany.PRL.2007}.  This letter investigates the effects of the grains and the GBs on the overall magnetization in V-doped In$_2$O$_3$ thin films.



In$_{2}$O$_{3}$ is a promising candidate material to exhibit strong GB magnetism because there is electron accumulation at the surfaces of pure In$_{2}$O$_{3}$ grains \cite{King.PRL.2008, King.PRB.2009}  and  it can also be grown strongly oxygen-deficient. There have been a number of studies of doped In$_{2}$O$_{3}$  following the initial identification of strong magnetism for Cr doping \cite{Philip.Nat.Mat.2006}.  FM has been seen for doping with Mn \cite{Jiang.APL.2010}, Fe \cite{Jiang.ASS.2009, Xu.APL.2009}, Co \cite{Hakimi.PRB.2011}, V \cite{Gupta.JAP.2007} and also  in undoped, but oxygen-deficient, In$_{2}$O$_{3}$  \cite{Hong.PSS.2007}  and as a surface effect in ITO \cite{Panguluri.PRB.2009, Xia.Physica.B.2011}. Vanadium is interesting and a good dopant for our study because V$^{3+}$ can substitute for In$^{3+}$, however its smaller ionic radius (78pm) compared with (94pm) for In$^{3+}$ will cause local strains and enhance the formation of GBs; it has been shown to be strongly ferromagnetic in In$_{2}$O$_{3}$ \cite{Gupta.JAP.2007} in spite of the fact that it does not form compounds with In and O that are ferromagnetic at room temperature.

In this letter we describe measurements on a series of V-doped In$_{2}$O$_{3}$  films that were fabricated such that they had different grain sizes.  In order to investigate the origin of the observed FM, we measured the structural, magnetic, electrical, optical and magneto-optical properties of V-doped In$_2$O$_3$ films. X-ray diffraction (XRD) measurements  allow us to estimate the preferred orientation, grain size, whereas the structural perfection around the TM site is studied by extended x-ray absorption fine structure (EXAFS). Magnetic properties were characterized by SQUID magnetometry which detects contributions from both spin and orbital magnetism  and by magnetic circular dichroism (MCD),  which is a powerful tool to confirm the energy of the magnetic states that give rise to the magnetism but since it is due to electric dipole transitions only detects orbital moments \cite{Neal.PRL.2006, Hakimi.APL.2010, Gehring.JMMM.2012}. We show that the overall magnetism originates from both the grains and the GBs in this material. The samples with smaller grain size have larger magnetization but a smaller MCD signal, whereas the samples with with a larger grain size have a smaller magnetization but larger MCD signal; this is due to the different contribution of the orbital magnetism in the two components.


\section{Experiment}

Polycrystalline thin films of (In$_{0.95}$V$_{0.05}$)$_2$O$_3$   were deposited on $c$-cut sapphire substrates at temperatures ranging between 300 and 550 $\,^{\circ}{\rm C}$ in 10$^{-5}$ Torr of oxygen by pulsed laser deposition (PLD)  with a XeCl laser (308 nm). The laser was used at a pulse repetition rate of 10 Hz, which gave an energy density up to 400 mJ/pulse. The target-to-substrate distance was 40 mm. Bulk targets for ablation were prepared by a standard solid-state reaction routine, as described elsewhere \cite{Hasan.2013}. Phase identification and structural properties of all the films were characterized by XRD  on a Bruker diffractometer in the $\theta$-2$\theta$ mode using Cu $K\alpha$ radiation ($\lambda$ = 1.5406 \AA). X-ray absorption fine structure (XAFS) measurements were performed at beamline 20-ID-B at the Advanced Photon Source using a microfocused beam incident on the sample at approximately 5$^{\circ}$ glancing angle. The magnetization measurements were performed in a SQUID magnetometer. The thickness of the films was measured with a Dektak profilometer and was kept almost constant at 120 $\pm$ 10 nm. Temperature dependence of the resistance  and Hall effect  measurements were performed in the Van der Pauw geometry, in a continuous-flow helium cryostat in the range 5 - 300 K in  fields up to 1 T. Optical absorption spectra and magneto-optical spectra were measured in Faraday geometry by using a photo-elastic modulator   from 2.0 eV to 3.5 eV in   fields up to 1.8 T. Since this energy range is below the band-gap of  In$_2$O$_3$, the magneto-optical response probes the magnetic polarization of any gap states.

\subsection{Structural Properties}

\begin{figure}
   \centering
        \includegraphics[width=0.7\textwidth]{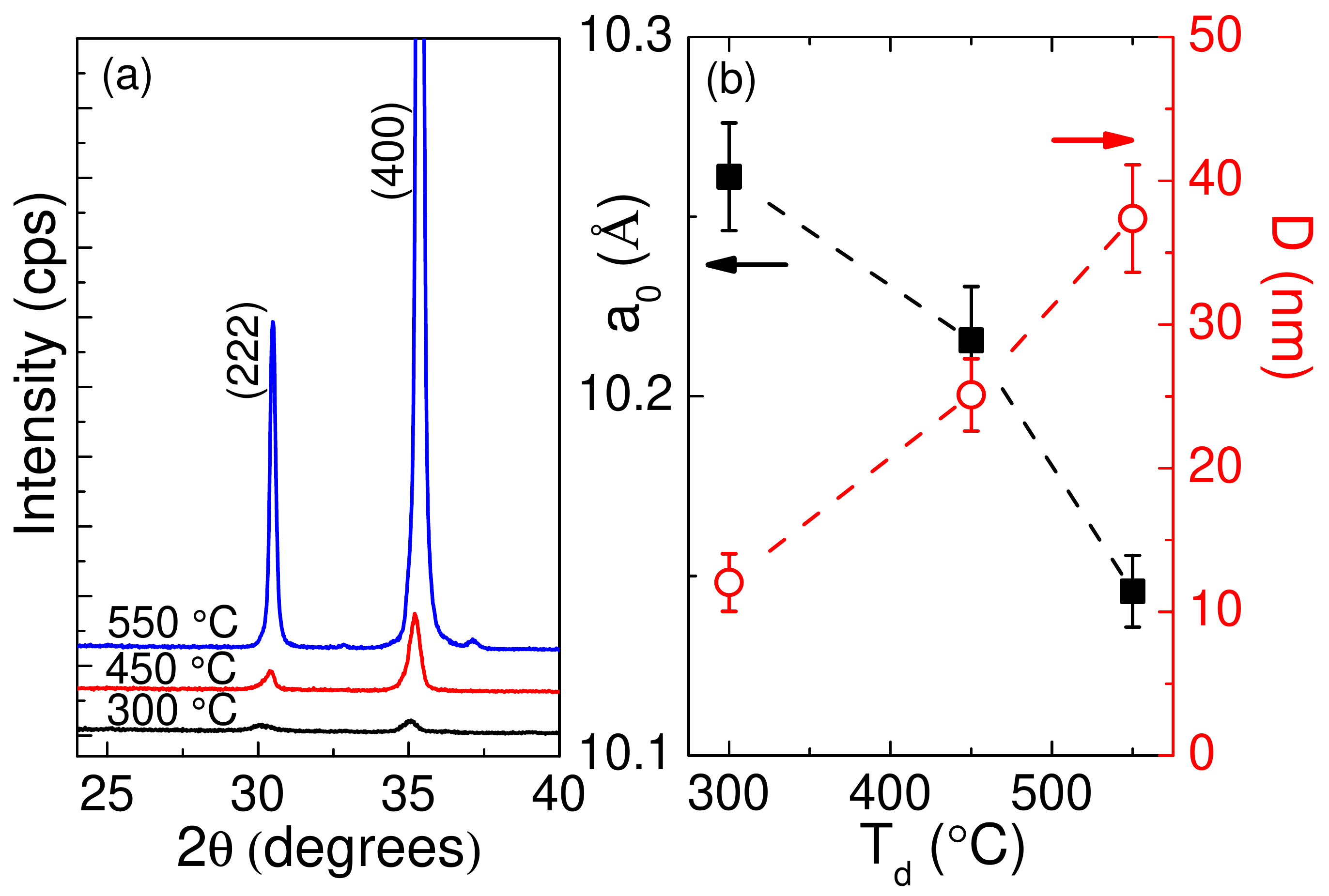} 
       \caption{\label{fig:fig_1} (Color Online) (a) $\theta-2\theta$ XRD diffractogram focusing on the In$_2$O$_3$ (222) and (400) reflection peaks for V-doped In$_2$O$_3$ films grown at different substrate temperatures. Displacement was applied for clarity.  (b) the lattice constant, $a_0$ (solid squares) and the grain size, $D$ (open circles), as a function of deposition temperature for V-doped In$_2$O$_3$ films. }
\end{figure}

Fig.\ref{fig:fig_1}(a) shows the $\theta-2\theta$ XRD scans of  (In$_{0.95}$V$_{0.05}$)$_2$O$_3$ films. All peaks were indexed assuming the same bixbyite cubic structure as pure In$_2$O$_3$. No evidence of any secondary phases of V was observed within the detection limit. As shown in Fig.\ref{fig:fig_1}(b), the lattice   constant was found to decrease as the deposition temperature increases, which is due to the smaller ionic radius of V$^{3+}$ (78 pm) compared with that of In$^{3+}$ (94 pm); this suggests that more V ions are incorporated into the host lattice as the deposition temperature increases.    The average size of the crystalline grains, $D$, in the films was estimated using the Debye-Scherrer method, $D = \frac{0.94 \lambda}{\beta \cos{\theta}}$, where $\lambda$ is the wavelength of the x-ray, $\beta$ the full width at half maximum  of (222) peak and $\theta$ the diffraction angle.  The grain size was found to increase with substrate temperature,  shown in Fig.\ref{fig:fig_1}(b),  from 12.5 to 34.4 nm for V-doped films.

\begin{figure}
   \centering
         \includegraphics[width=0.7\textwidth]{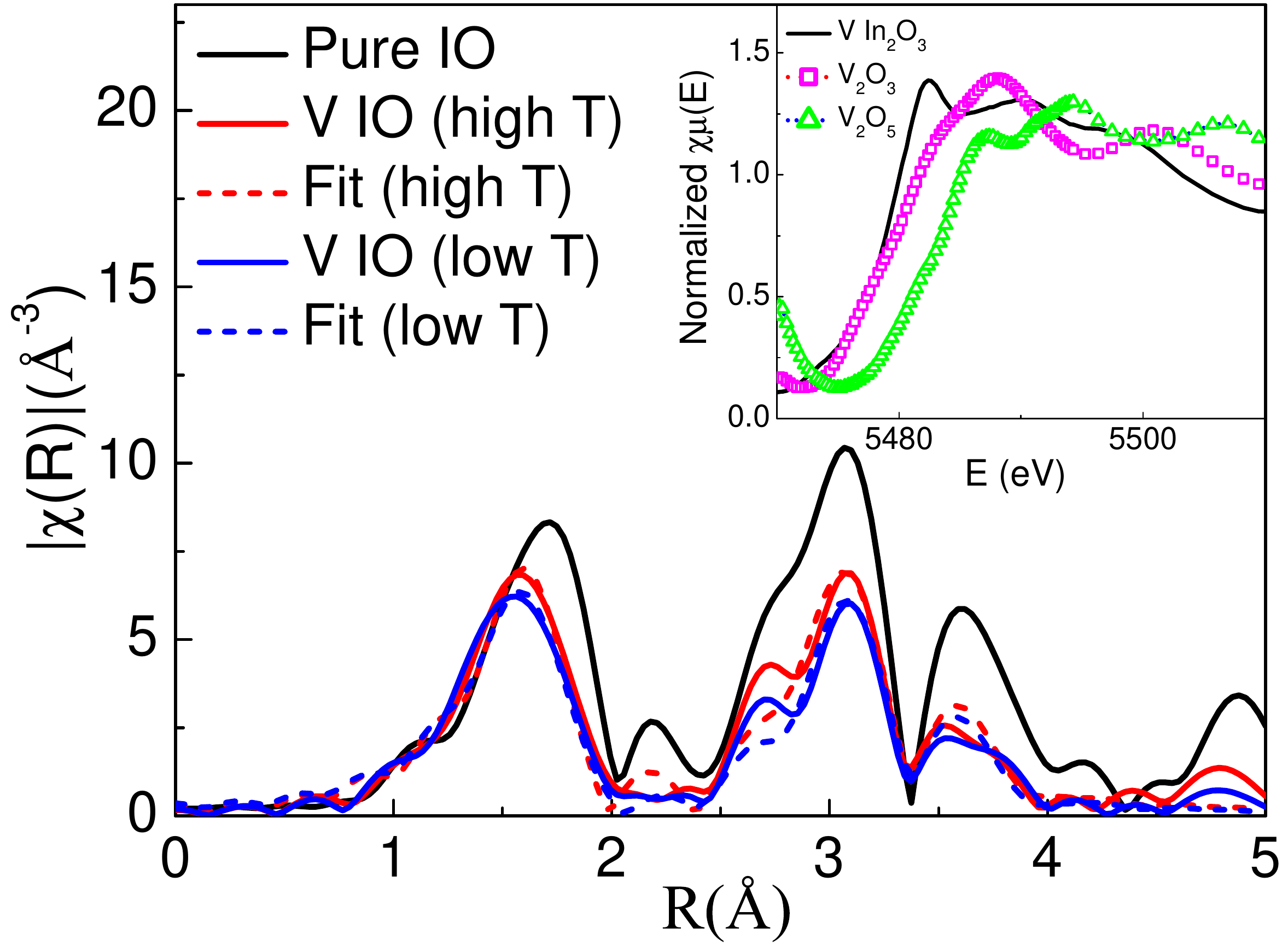} %
       \caption{\label{fig:fig_2} (Color Online) $k^3$ weighted Fourier transforms of the EXAFS data for pure In$_2$O$_3$ film (black line) and two V-doped In$_2$O$_3$ films grown at 550 and 300  $\,^{\circ}{\rm C}$ (red and blue lines), respectively.   The transform range was $k=2-11.5$ \AA$^{-1}$. A two-site model based in the In$_2$O$_3$ bixbyite structure was used. The fitting was denoted by dashed lines. Inset: normalized near-edge XAFS spectra for V-doped In$_2$O$_3$ grown at 300 $\,^{\circ}{\rm C}$ (black line) and different V-oxides, V$_2$O$_3$ and V$_2$O$_5$ (pink squares and green triangles). }
\end{figure}

XAFS measurements were used to look for potential second phase formation. An analysis was carried out that was very similar to that described  earlier \cite{Heald.JPCS.2012}.  As seen in the inset in Fig.\ref{fig:fig_2}, the position of the near edge of V clearly indicates that all of the V is in the 3+ valence state.  EXAFS is used to explore the local environment of the doped TM ions. Fig.\ref{fig:fig_2}  shows the Fourier transformed data for  pure In$_2$O$_3$ and two V-doped In$_2$O$_3$ films, which were grown at 300 and 550  $\,^{\circ}{\rm C}$. As shown  earlier \cite{Heald.JPCS.2012}, the dopant sites in the In$_2$O$_3$ lattice can have widely varying degrees of disorder. The V data can be well-fitted, suggesting that V occupies substitutional sites, as shown in Fig.\ref{fig:fig_2}.  To fit the data, we initially used a  two-site model \cite{Heald.JPCS.2012}, which gave a good fit to all V data, although there is a mismatch near $R=2.6$ \AA. The fit could not be significantly improved by the addition of secondary phases, such as metal V, VO, VO$_2$, V$_2$O$_3$. Therefore, we suggest that this mismatch may be due to   distortion around the V site, such as  GBs. Reduced disorder in the data for the film grown at 550 $\,^{\circ}{\rm C}$ suggested that the crystallinity in the neighborhood of the V ions was improved by growth at a higher temperature as found by XRD.

\subsection{Optical Properties}

\begin{figure}
   \centering
         \includegraphics[width=0.7\textwidth]{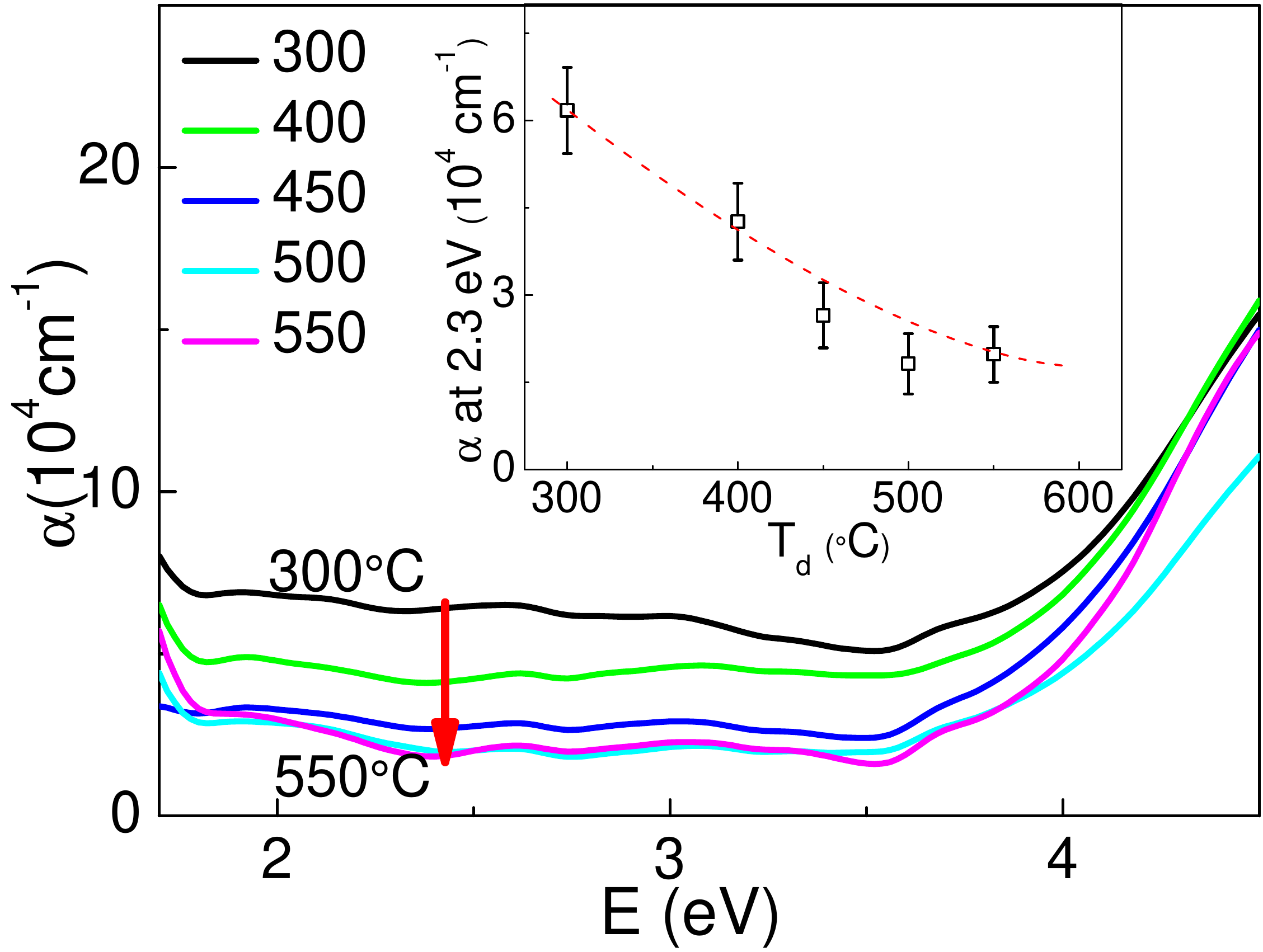}
       \caption{\label{fig:fig_3} (Color Online) Room temperature optical loss spectra for 5 at.\% V-doped In$_2$O$_3$ films, deposited at substrate temperatures ranging from 300 to 550 $\,^{\circ}{\rm C}$. Inset:  room temperature loss coefficient obtained at 2.3 eV for V-doped In$_2$O$_3$ films as a function of substrate temperatures.}
\end{figure}

Fig.\ref{fig:fig_3} shows the optical absorption, together with  the surface scattering measurements at room temperature  as a function of photon energy for V-doped In$_2$O$_3$ films grown at different temperatures. The transmittance below the band gap is found to increase with increase in substrate temperature. The absorption in the low energy region   decreases  significantly with increasing substrate temperature,  as   shown in the inset in Fig.\ref{fig:fig_3}.  Thus,  the optical data support the EXAFS and the conclusions drawn from the XRD that, as deposition temperature increases, the crystallinity improves and grain size increases. Similar behavior was observed for Mn- and Fe-doped In$_2$O$_3$ films \cite{Qi.APL.Mat.2013}.

\subsection{Electrical Properties}

\begin{figure}
   \centering
         \includegraphics[width=0.7\textwidth]{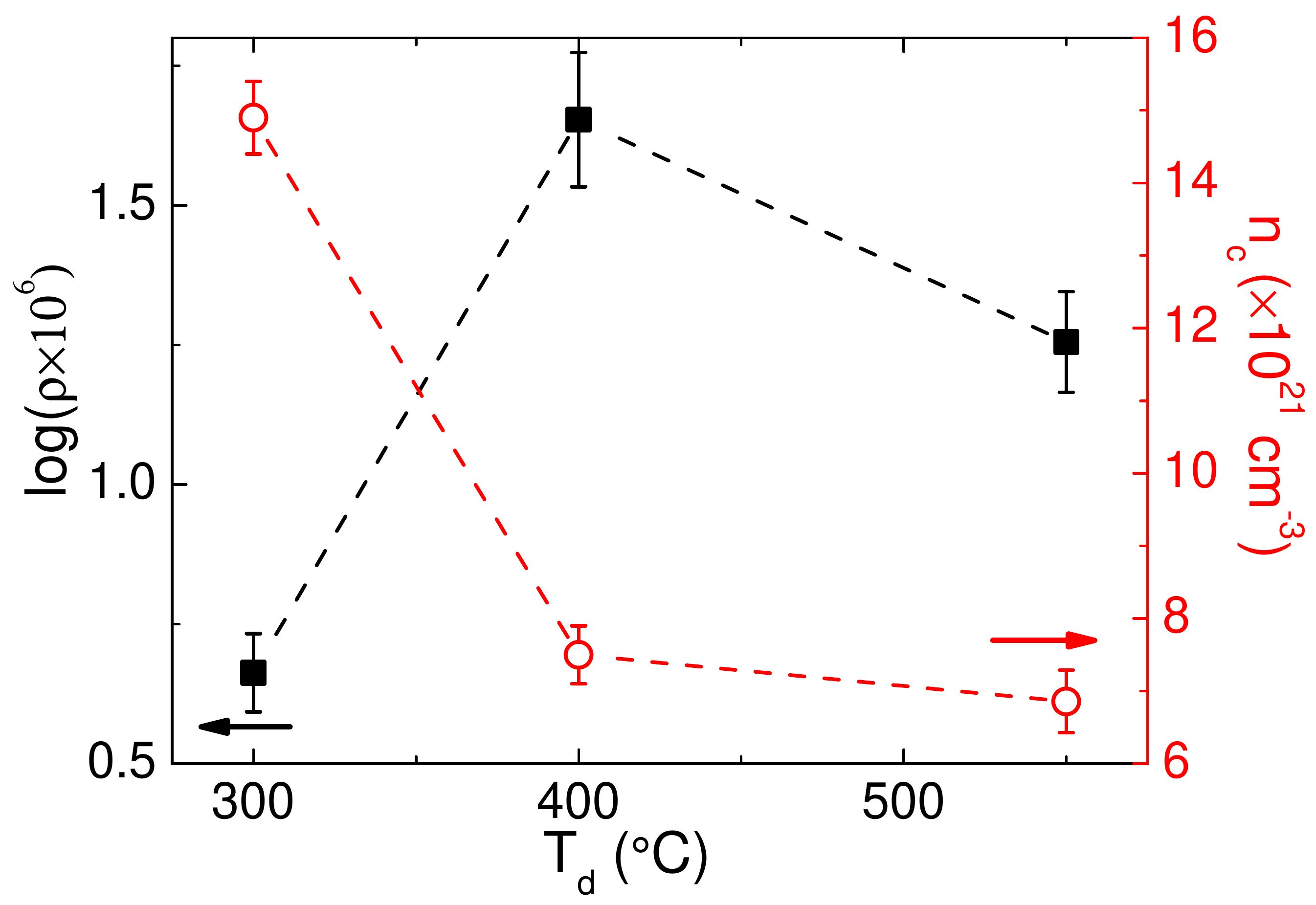}
       \caption{\label{fig:fig_4} (Color Online) Resistivity, $\rho$ (solid squares) and carrier density, $n_c$ (open circles), obtained at 300 K, as a function of deposition temperature for V-doped films. The resistivity of the film grown at 300 $\,^{\circ}{\rm C}$ was 4.60 $\times$ 10$^{-6}$ $\Omega$m. }
\end{figure}

Measurements of the Hall effect and the resistivity were performed in Van der Pauw geometry at both 5 and 300 K to explore the carrier concentration, and resistivity. The films show metallic behaviour over the temperature range 5$<$T$<$300 K. The room temperature resistivity, $\rho$, and the carrier concentration, $n_c$,  for V-doped In$_2$O$_3$ films as a function of  deposition temperature for V-doped In$_2$O$_3$ films  are  presented in Fig.\ref{fig:fig_4}.    The carrier concentration was comparable to that found for undoped In$_2$O$_3$ films grown at low oxygen pressure. Further experiments will be required to investigate if the electrons that are localized  at the GBs carry an appreciable fraction of the total current in the material.

\subsection{Magnetic Properties}

\begin{figure}[ht]
   \centering
         \includegraphics[width=0.7\textwidth]{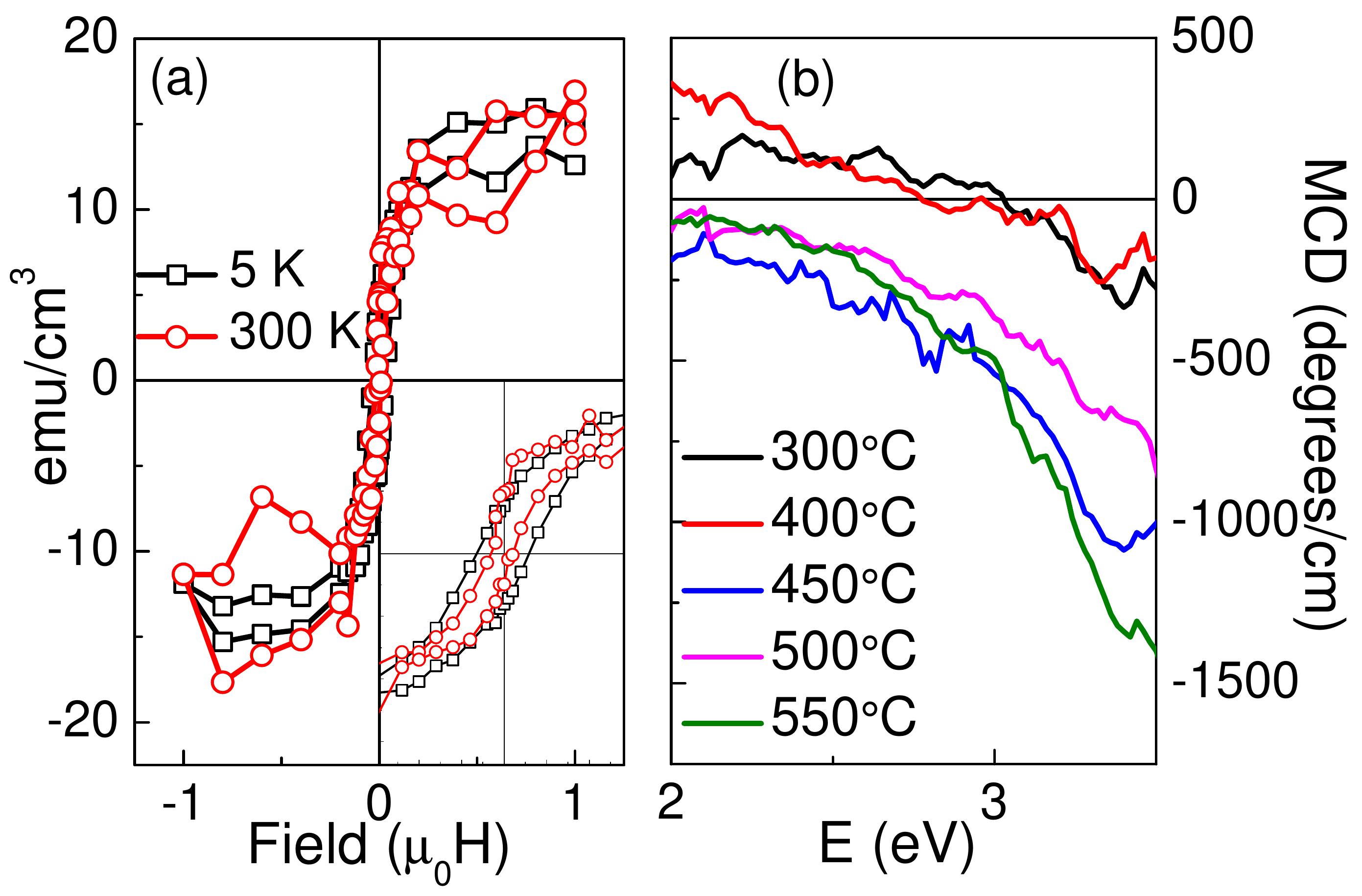}
       \caption{\label{fig:fig_5} (Color Online) (a) hysteresis loops at 5 and 300 K for 5 at.\% V-doped In$_2$O$_3$ film deposited at 300 $\,^{\circ}{\rm C}$. Inset: loops at low fields. (b) room temperature MCD data as a function of photon energy for V-doped In$_2$O$_3$ films deposited at various temperatures.}
\end{figure}

\begin{figure}[ht]
   \centering
         \includegraphics[width=0.7\textwidth]{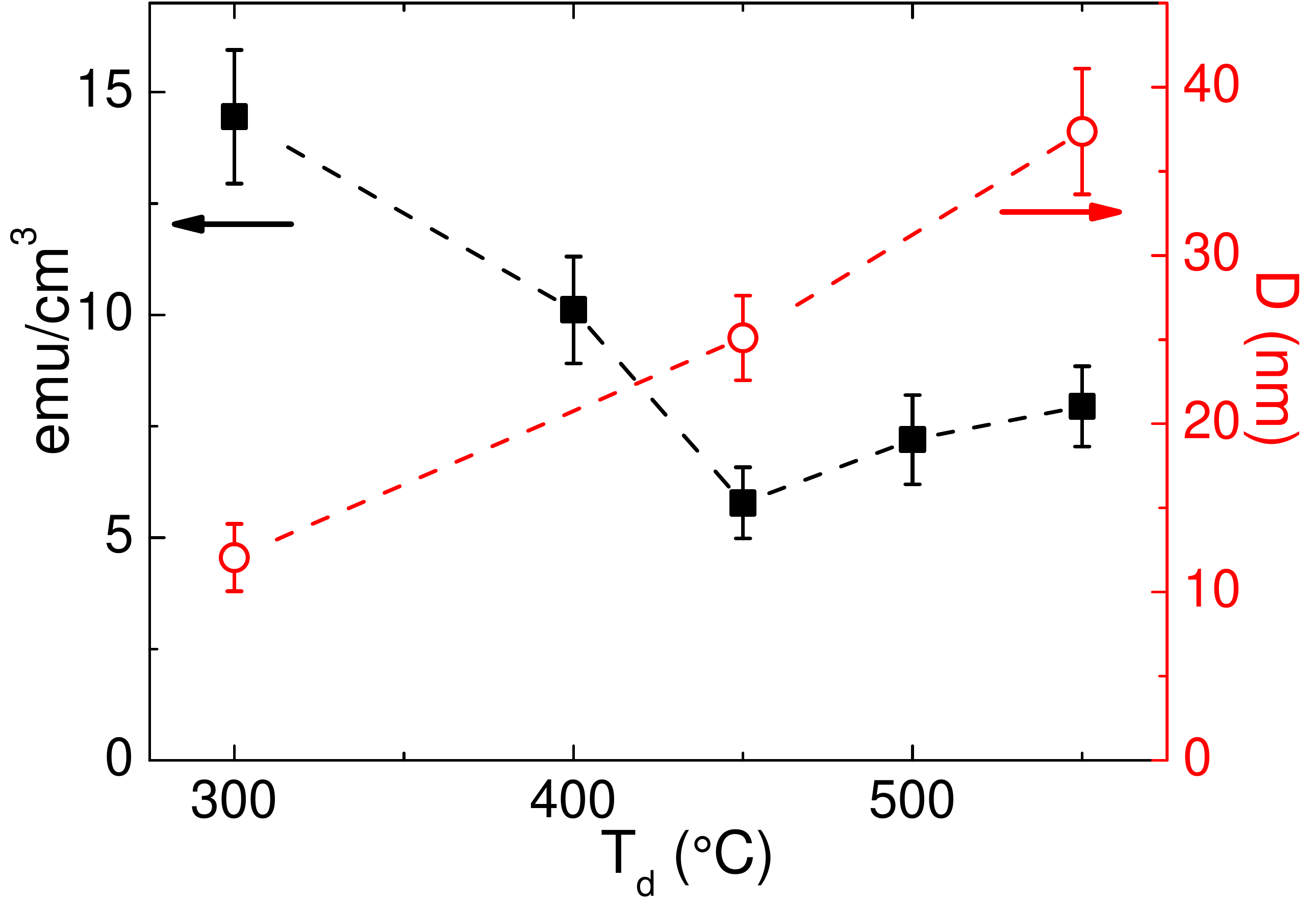}
       \caption{\label{fig:fig_6} (Color Online) The saturation magnetization (solid squares), $M_s$, and the grain size (open circles), $D$, as a function of deposition temperature for V-doped In$_2$O$_3$ thin films.}
\end{figure}
\begin{table}
\centering
 \caption{\label{tab:tab_1} Thickness, $d$, grain size, $D$, loss coefficient, $\alpha$ at 2.3 eV, MCD at 3.3 eV, and saturation magnetization, $M_s$, in  units of both emu and emu/cm$^{3}$, for In$_2$O$_3$ doped with 5\% V  samples prepared at different substrate temperature, $T_d$.}
\begin{tabular}{p{0.8cm}|p{1.4cm}|p{1.9cm}|p{2.2cm}|p{2.7cm}|p{2.3cm}|p{2cm}}
  \hline
  $T_d$ & $d$  & D & $\alpha$ at 2.3 eV  & MCD at 3.3 eV  & \multicolumn{2}{c}{$M_s$} \\ \cline{6-7}
  ($^{\circ}{\rm C}$) & (nm) &(nm)& (10$^4$ cm$^{-1}$)&(degrees/cm)& ($\times 10^{-5}$ emu)& (emu/cm$^{3}$) \\ \hline
  300 & 115 $\pm$ 5 & 12.1 $\pm$ 0.1  &6.2 $\pm$ 0.3 & 220 $\pm$ 25  & 3.2 $\pm$ 0.3 & 14.5 $\pm$ 1.3\\
  400 & 115 $\pm$ 5 & -               &4.3 $\pm$ 0.2 & 230 $\pm$ 25  & 2.3 $\pm$ 0.2 & 10.1 $\pm$ 0.9\\
  450 & 125 $\pm$ 5 & 25.1 $\pm$ 0.2  &2.7 $\pm$ 0.1 & 980 $\pm$ 75  & 1.4 $\pm$ 0.1 & 5.8 $\pm$ 0.5  \\
  500 & 125 $\pm$ 5 & -               &1.8 $\pm$ 0.1 & 660 $\pm$ 65  & 1.8 $\pm$ 0.1 & 7.2 $\pm$ 0.7 \\
  550 & 110 $\pm$ 5 & 37.4 $\pm$ 0.2  &2.0 $\pm$ 0.1 & 1130$\pm$ 80  & 1.6 $\pm$ 0.1 & 7.9 $\pm$ 0.2  \\
  \hline
\end{tabular}
 \end{table}

Magnetic properties were investigated by SQUID magnetometry,  the data shown in this letter has been analyzed by subtracting the diamagnetic contribution of the pure sapphire substrate. Fig.\ref{fig:fig_5}(a) shows the in-plane M-H curves of V-doped In$_2$O$_3$ films at 5 and 300 K. The temperature independent magnetization in this case and all the other TM-doped In$_2$O$_3$ samples agrees with that found in Fe-doped In$_2$O$_3$ films by   Jiang {\it et al} \cite{Jiang.JAP.2011} and Park {\it et al} \cite{Park.APL.2012}. They found that for Fe-doped In$_2$O$_3$ thin films, magnetization was temperature independent in a metallic sample, whereas  temperature dependent magnetization was observed in a semiconducting sample. Fig.\ref{fig:fig_5}(b) shows the room temperature MCD as a function of photon energy for V-doped In$_2$O$_3$ films.  The MCD spectra, together with  the saturation magnetization, was unchanged down to 10 K indicating a lack of signal from any V minority phases.


Fig.\ref{fig:fig_6} shows the saturation magnetization and the grain size as a function of deposition temperature for V-doped In$_2$O$_3$ thin films. It is clear that as the deposition temperature increases, the grain size increases, whereas the saturation magnetization decreases. Information of the thickness, $d$, grain size, $D$, loss coefficient, $\alpha$,   saturation magnetization, $M_s$, and MCD for samples prepared at different substrate temperature, $T_d$, is summarized in table~\ref{tab:tab_1}. The saturation magnetization was analyzed in terms of the thickness and the surface area of the sample; surface area is almost constant during the PLD process, about 20 mm$^2$ and the thickness of samples was kept almost constant at 120 nm. In Ref~\cite{Panguluri.PRB.2009}, it was found that undoped In$_2$O$_3$ had a moment of 0.5 emu/cm$^3$ which is small compared with the values  found here 2-15 emu/cm$^3$. It is important to note that our results are significantly large compared also with other published results which means that our values are more robust \cite{Hong.PSS.2007, Philip.Nat.Mat.2006, Panguluri.PRB.2009}.

\begin{figure}[ht]
   \centering
         \includegraphics[width=0.8\textwidth]{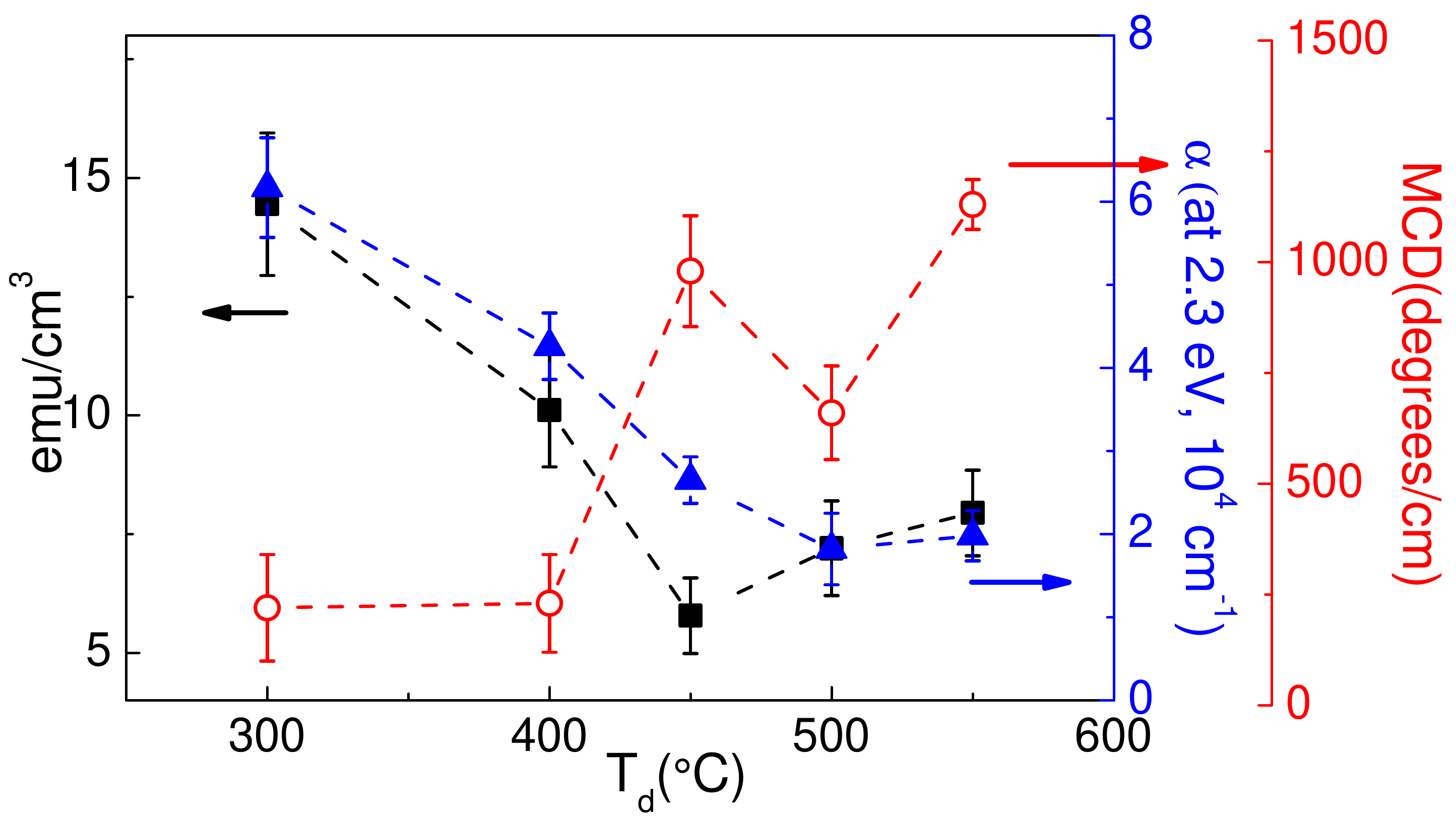} 
       \caption{\label{fig:fig_7} (Color Online) The variation of the saturation magnetization, $M_s$ (solid squares), room temperature MCD at 3.3 eV (open circles) and  absorption coefficient at 2.3 eV (solid triangles) for 5 at.\% V-doped In$_2$O$_3$ films, as a function of deposition temperatures ranging between 300 and 550 $\,^{\circ}{\rm C}$.}
\end{figure}

Fig.\ref{fig:fig_7} shows the variation of the room temperature saturation magnetization,  MCD and optical absorption coefficient for V-doped films, as a function of substrate temperature.    The saturation magnetization decreases  as a function of deposition temperature, behaving similarly to the optical absorption. This shows that when the grain size gets larger at higher deposition temperature, the scattering from the sample surface becomes weaker, thus leading to smaller absorption; this is in consistent with measurements of D found from XRD.  More grain boundaries form with smaller grain size in the normalized volume. The decrease of saturation magnetization as the grain size increases strongly suggests that much of the observed magnetism originates from the grain boundaries.

The  MCD values quoted in Fig.7 were measured at 3.3 eV, which is just below the band-gap; this is believed to correspond to transitions to spin-polarized gap states. The MCD signal would be zero if the magnetization were  zero, but they are not necessarily proportional because of the importance of spin-orbit coupling. The MCD spectra shown in Fig.\ref{fig:fig_5}(b) peak at the band edge as expected for an oxide but actually fall into two distinct groups. There is a small dispersive signal for the samples grown at 300 and 400 $\,^{\circ}{\rm C}$, where the samples have a high magnetization, and a larger absorption-like signal for the higher temperatures where the magnetization is reduced. These spectra are similar to those found for Mn and Fe-doped In$_2$O$_3$ \cite{Qi.APL.Mat.2013}. The Mn-doped sample had a high magnetization  when grown at low temperatures and the MCD spectrum was dispersive. The maximum magnetization for the Fe-doped sample occurred for high growth temperature and this was accompanied by an absorptive MCD spectrum. The magnitude of the absorptive MCD spectrum for Fe was much larger than that for the dispersive signal for Mn, even though the magnetization was significantly higher for Mn.

The MCD depends on a change in the angular momentum in the optical transitions and this is related to the observed magnetization via the spin-orbit coupling. For any one material the MCD follows the magnetization as a function of temperature or magnetic field. However, for different materials,   the magnitude of the MCD also depends on the spin-orbit coupling that is necessary to induce an orbital moment. For the purposes of this discussion boundary magnetism and grain magnetism are two different quantities.  Thus it appears that the magnetism located at grain boundaries arises from electrons with rather weak spin-orbit coupling  hence a large grain boundary magnetism gives a small MCD signal. On the other hand the electrons that cause the magnetism in the grains have a much larger spin-orbit interaction and hence a large MCD signal may be generated by a smaller magnetization.  These results are strong evidence to support the contention that the observed FM  originates from both the grain boundaries and the grains. We note that the magnetization from the grains increases when the grains form a larger fraction of the whole sample and also the concentration of V ions has increased as indicated by the change in lattice constant   observed by XRD.

The large $M_s$ from the boundaries has a relatively low orbital contribution and hence gives a small MCD signal. As the grains get larger the grain contribution begins to be more important and this has a larger orbital contribution and hence gives a relatively large MCD. The GB FM, however, dominates in the samples.





\section{Conclusion}

In conclusion, we have observed room temperature FM  in V-doped In$_2$O$_3$ films by SQUID and MCD measurements. The increase in grain size with deposition temperature was observed in doped films, similar to that found in Mn- and Fe-doped films \cite{Qi.APL.Mat.2013}. Structural characterization indicated V ions are substituted at the In sites in the In$_2$O$_3$ host,  and more V ions were incorporated into the host lattice at higher deposition temperature. Although the fit to the EXAFS data shows some mismatch, the existence of secondary phases of V was excluded. The contribution from secondary phases in doped films was also excluded  by low temperature MCD measurements.  The large magnetization seen for V-doped samples, when the grains were small  indicates that GB FM is observed in V-doped  In$_2$O$_3$ films, similarly  to that found in Mn-doped ZnO and In$_2$O$_3$ films \cite{Straumal.PRB.2009, Qi.APL.Mat.2013}, as a growing evidence of GB  model. The magnetization of V-doped samples is relatively large compared with other oxide materials - it may be significant that unlike Mn, Co which are present as divalent ions and Fe where magnetization depends on having some divalent Fe \cite{Jiang.JAP.2011} as well as trivalent Fe the V ions are all V$^{3+}$ and yet the magnetization is unusually large. The system with Gb FM  would be a promising source of spin-polarized currents if the   electrons that are polarized at the GBs  carry an appreciable fraction of the total current in the material. Therefore further investigation is required.


\section*{Acknowledgement}
Use of the Advanced Photon Source, an Office of Science User Facility operated for the U.S. Department of Energy (DOE) Office of Science by Argonne National Laboratory, was also supported by the U.S. DOE under Contract No. DE-AC02-06CH11357. The work is also financially supported by NSFC (51025101). 


\end{document}